\documentclass[fleqn,10pt]{wlscirep}
\title{A norm knockout method on indirect reciprocity to reveal indispensable norms\footnote{The final version was published in {\it Scientific Rreports}. How to cite this article: Yamamoto, H., Okada, I., Uchida, S., \& Sasaki, T. A norm knockout method on indirect reciprocity to reveal indispensable norms. {\it Sci. Rep.} 7, 44146; doi: 10.1038/srep44146 (2017).}}

\author[1,*]{Hitoshi Yamamoto}
\author[2,+]{Isamu Okada}
\author[3,+]{Satoshi Uchida}
\author[4,+]{Tatsuya Sasaki}
\affil[1]{Rissho University, Department of Business Administration, Tokyo, 141-8602, Japan}
\affil[2]{Soka University, Department of Business Administration, Tokyo, 192-8577, Japan}
\affil[3]{RINRI Institute, Research Center for Ethiculture Studies, Tokyo, 102-8561, Japan}
\affil[4]{University of Vienna, Faculty of Mathematics, Vienna, 1090, Austria}

\affil[*]{hitoshi@ris.ac.jp}
\affil[+]{these authors contributed equally to this work}

\begin{abstract}
Although various norms for reciprocity-based cooperation have been suggested that are evolutionarily stable against invasion from free riders, the process of alternation of norms and the role of diversified norms remain unclear in the evolution of cooperation.
We clarify the co-evolutionary dynamics of norms and cooperation in indirect reciprocity and also identify the indispensable norms for the evolution of cooperation.
Inspired by the gene knockout method, a genetic engineering technique, we developed the norm knockout method and clarified the norms necessary for the establishment of cooperation.
The results of numerical investigations revealed that the majority of norms gradually transitioned to tolerant norms after defectors are eliminated by strict norms.
Furthermore, no cooperation emerges when specific norms that are intolerant to defectors are knocked out.
\end{abstract}
\begin{document}

\flushbottom
\maketitle
%
%
\thispagestyle{empty}

\section*{Introduction}
Reciprocity is a fundamental mechanism that underlies all cooperative societies. Theoretically it is well known that direct reciprocity, typified by the “I’ll help you if you help me” attitude, promotes cooperative regimes \cite{Trivers1971,Axelrod1981}. However, in recent societies that have high relational mobility, indirect reciprocity such as “I’ll help you and somebody else will help me” plays a more important role in promoting cooperation. Indirect reciprocity has therefore been the focus of much research in the interdisciplinary fields in recent decades  \cite{Alexander1987,Sugden1986,kandori1992,wedekind2000,Panchanathan2004}.

Many theoretical studies on indirect reciprocity have explored norms that become evolutionarily stable against defection and the invasion of free riders, and several typical norms have been proposed \cite{Ohtsuki2004a,Nowak2005,Ohtsuki2006,Takahashi2006}. These approaches have clarified the robust norms that can maintain the cooperative regime.
The norms in the studies on the indirect reciprocity are regarded as assessment rules that label the other's action as either Good or Bad.
They include tolerant norms that assess cooperative behaviors toward defectors as good\cite{Sugden1986} and strict norms that assess such behaviors as bad\cite{pacheco2006}. Other theoretical studies analysing the global dynamics of norms assume that at most a few robust norms are shared in the population \cite{Ohtsuki2007,uchida2010competition,uchida2010effect}.

Their approaches have clarified the robustness of the norms against invasion of other norms including free riders when the norms are acceptable in the population.
However, little is known about a process by which gradual changes toward cooperation occur as new norms emerge and compete, which is to say, the co-evolutionary process of norm-diversity and cooperation.
A study on the indirect reciprocity has dealt with co-existing different norms and has analysed their frequencies in the population as a consequence of a dynamical process\cite{Brandt2004}.
In the study, each individual keeps a private image of everyone else and errors of perception and implementation are included in a limited strategy space.
Although they have considered some action rules and assessment rules, all possible norms in indirect reciprocity have not been studied all-together.
How cooperation evolves cannot be fully understood unless the evolution of norms is also considered.

It is thus a challenging task to theoretically understand how cooperation can be formed even under a collection of norms in a social system.
How is the co-existence of cooperation and diversity possible at all?
Are there any indispensable norms needed to facilitate the evolution of cooperation in the melting pot of norms, even though some norms never become dominant? Do norms that could be accepted as a result of the co-evolutionary process have common aspects? These questions can be addressed only if all possible norms are considered, and the combination of norms governing a group can evolve.

Here we explore the dynamics of co-evolution of cooperation by using different social norms.
The process of the evolution of norms has a transition from stricter to tolerant norms.
Additionally, we find a set of norms that seem not to have an impact on promoting cooperation, but are fundamental to allow a transition to a cooperative regime from a defective regime.

\section*{Results}

\begin{figure}[ht]
\centering
\includegraphics[width=\linewidth]{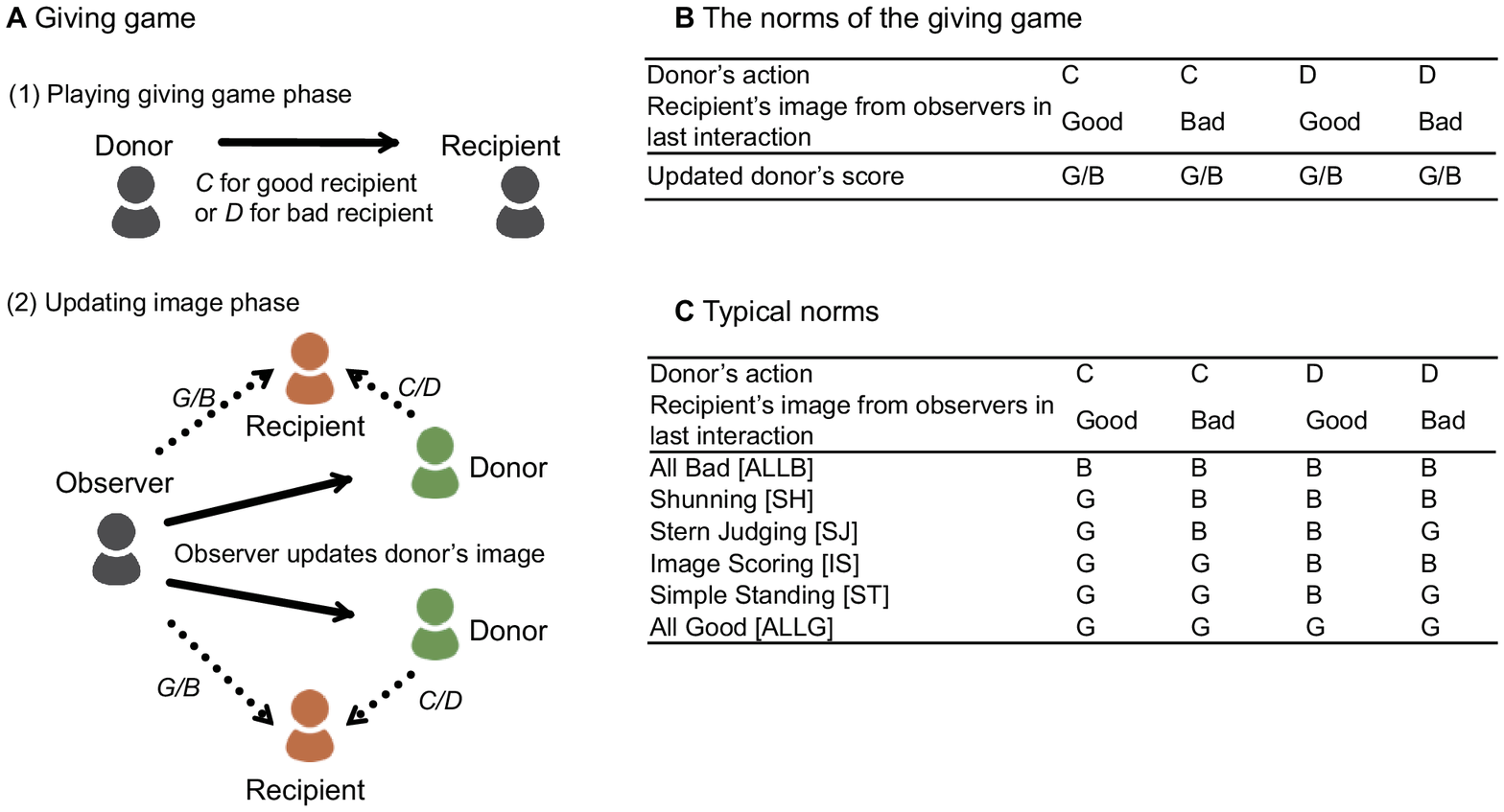}
\caption{{\bf The norms of cooperation and simulation framework}. {\bf \sf a}, (1) If the donor’s image of a recipient is Good, the donor gives the recipient something with personal cost $c$ and the recipient receives benefit $b$. Nothing happens otherwise. (2) In the Updating image phase, the observer updates the evaluation to the donor on the basis of the donor’s action (Cooperation [C] / Defection [D]) and the observer’s evaluation (Good [G] / Bad [B]) of the recipient. {\bf \sf b}, Each agent adopts an evaluation rule of the donor that depends on the donor’s action and the recipient’s image. This combination of Good/Bad is the norm held by the agent. There are a total of $2^4 = 16$ possible norms.
In this phase, each agent evaluates and updates its image to all donors.
{\bf \sf c}, Typical norms can be expressed in the manner shown in this table. Typical norms include Shunning [SH] = GBBB, Stern Judging [SJ] = GBBG, Image Scoring [IS] = GGBB, and Simple Standing [ST] = GGBG.
SH is a strict norm where any action for a Bad recipient is assessed as Bad.
ST is a tolerant norm where any action for a Bad recipient is assessed as Good.
SJ is an intermediately strict norm where cooperation for a Bad recipient is assessed as Bad while defection is Good.
In contrast, IS does not use an image to recipient but uses only donor's action.
If the donor's previous action is C, then IS evaluates the donor as Good, otherwise IS evaluates the donor as Bad. 
}
\label{fig:framework}
\end{figure}

\begin{figure}[ht]
\centering
\includegraphics[width=\linewidth]{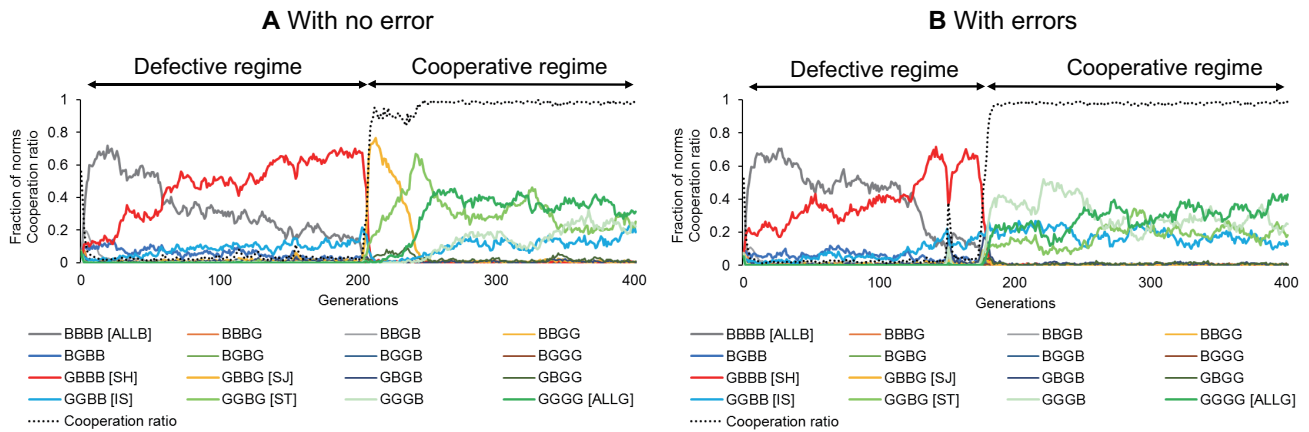}
\caption{{\bf Time series of typical simulation runs with all norms. With no error (left panel) and with errors (right panel).}
{\bf \sf a}, The average frequencies of 16 norms and the cooperation of the overall society. The black dotted line is the cooperation ratio. Parameters: $b = 5, c = 1, N = 500, R = 500, G = 1000, p = 0, q = 0$.
When SH and ALLB coexist, cooperation does not emerge. When ALLB is completely driven out by SH, SJ invades and the cooperation ratio abruptly rises. At the same time, SH is driven out by SJ.
After cooperation is completely achieved, SJ permits the invasion of ST, and also coexists with other tolerant norms (IS, ST, GGGB, and ALLG).
Finally, strategies whose norm is expressed as GG** (in other words, norms that constantly cooperate if cooperation has been selected in the past by the recipient) coexist.
In {\bf \sf b}, both errors in perception and implementation were introduced, and simulation similar to {\bf \sf a} was run ($b = 5, c = 1, N = 500, R = 500, G = 1000, p = 0.001, q = 0.001$). 
As in {\bf \sf a}, when SH and ALLB coexist, cooperation does not emerge.
However, cooperation is achieved without going through SJ.}
\label{fig:alternation1}
\end{figure}

\begin{figure}[ht]
\centering
\includegraphics[width=\linewidth]{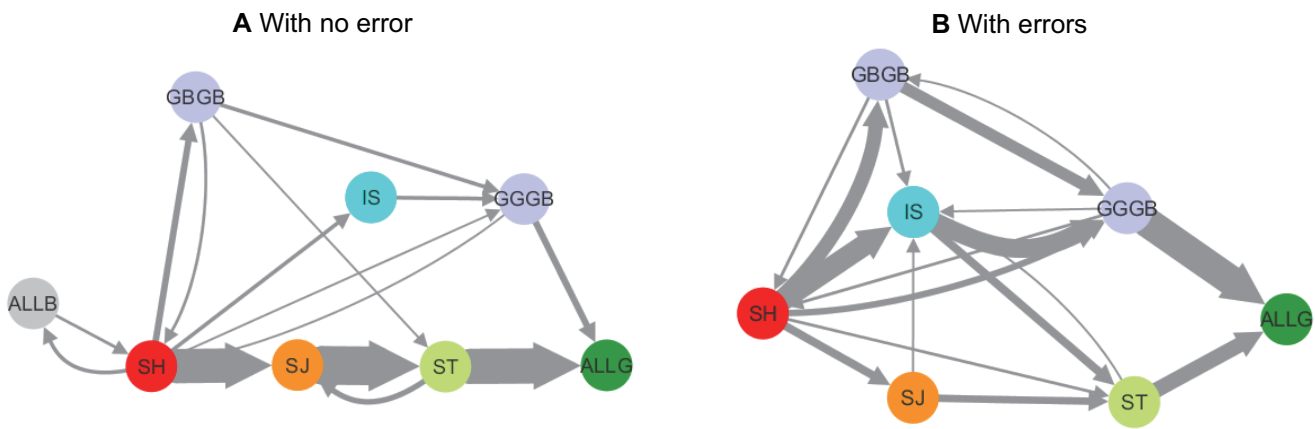}
\caption{{\bf The alternation patterns of the majority of norms with 50 replications. With no error (left panel) and with errors (right panel).}
{\bf \sf a}, The panel shows the transition of norms with the greatest populations in a round of 20 generations, before the cooperation ratio exceeds 0.8, and 100 generations, after the cooperation ratio exceeds 0.8 (for a total of 120 generations).
For the sake of visibility, in a replication, we stop calculation when ALLG becomes the majority norm.
This is because in a state in which tolerant norms coexist, the norms with the greatest population frequently change place.
The thickness of the arrows corresponds to the number of times alternation of norms occurred. (See the Supplementary Information for details.)
The alternation of norms SH $\to$ SJ $\to$ ST $\to$ ALLG was observed to be stable.
In {\bf \sf b} , both errors in perception and implementation were introduced, and simulations similar to {\bf \sf a} were run ($b = 5, c = 1, N = 500, R = 500, G = 1000, p = 0.001, q = 0.001$).
As shown in {\bf \sf b}, the transition of majority norms is not distinct compared to the times when there were no errors.}
\label{fig:alternation2}
\end{figure}

Agent-based simulations \cite{Nigel1999,Roberts2008} are an optimal tool to tackle the challenge outlined above.
See Methods for the details of our agent-based model described by the ODD protocol\cite{Grimm2010}.
Using an evolutionary game theoretical framework and constructing an interaction model based on players’ private rules and local information, we model a giving game to elucidate the dynamics of the evolution of cooperation amid the coexistence of diverse norms (Fig. \ref{fig:framework}).
We conducted numerical simulations of all 16 possible norm combinations that could react to the four combinations of assessment criteria to clarify the dynamics of the evolution of cooperation from the melting pot of diverse norms.
Figure \ref{fig:alternation1}a shows time-series graphs of each norm’s population and cooperation ratio.
As shown, the majority undergo an alternation from strict to tolerant norms, mostly in the order of $\rm{SH \rightarrow SJ \rightarrow ST}$.
Figure \ref{fig:alternation2}a shows the transition in the norm with the greatest population ratio.
In many cases, the majority transitioned from the state where strict SH \cite{Takahashi2006} was the majority to SJ \cite{kandori1992,pacheco2006}.
Afterwards, the majority norm changed to tolerant ST \cite{Sugden1986,leimar2001,Panchanathan2003} and ALLG.
In contrast, as shown in Figs. \ref{fig:alternation1}b and \ref{fig:alternation2}b, in an environment with errors, alternation from strict norms to tolerant norms was observed.
However, the likelihood of going through SJ decreased.
Alternation paths through IS \cite{nowak1998,Nowak1998b}, which could not be seen in an environment without errors \cite{Lotem99}, increased.
It is important to note here that similar paths toward cooperation are observed when only ALLB-individuals are initially assumed.
New norms are created during the evolutionary process at the same time cooperation evolves. This indicates that cooperation and diversity of norms jointly evolve in the model.

Why does the alternation of norms emerge? For one thing, in states in which defection is dominant, ALLB (BBBB) and SH (GBBB) coexist and jointly form the majority. However, BGBB and IS (GGBB) continue to exist as the minority. The characteristic of these groups is having the evaluation rule of **BB. Evaluation rule **BB assesses donors that took D as B, regardless of the evaluation of the recipient. In states in which defection is dominant, those who adopt **BB strategies consider many partners as B. As a result, cooperation does not occur for the most part. The ALLB and SH norms thus survive because they do not lower their own cost. On the other hand, after cooperation is achieved, ALLG (GGGG), ST (GGBG), IS (GGBB), and GGGB coexist. The common characteristic of these norms is having the evaluation rule of GG**. Thus, reciprocally cooperating norms survive. Because SJ (GBBG), which becomes the majority temporarily when the cooperation ratio rises in an environment without errors, does not belong to either group, it cannot stably exist. Also, it is rare that SJ makes up the majority temporarily in an environment with errors. Meanwhile, because IS belongs to both norm groups with **BB and GG**, IS can constantly exist.

We discover several norms that are indispensable to the evolution of cooperation. Reputation-based cooperation cannot emerge without indispensable norms. To elucidate indispensable norms for the evolution of cooperation, we propose a novel analysis using the norm knockout method. This method enables us to determine which norms are indispensable for the evolution of cooperation. The norm knockout method is inspired by the targeted gene knockout technique used in genetic engineering \cite{Strepp1998}. Gene knockout, a genetic technique in which one of an organism’s genes is made inoperative, is used to research genes whose sequences are known but whose functions are not well-understood. Researchers infer the gene’s function from differences between the knockout animal and a normal animal. For simulating evolution, we utilized a method that removed only one particular norm from the population to understand whether that norm is an indispensable one that plays a critical role in the evolution of cooperation.

\begin{figure}[ht]
\centering
\includegraphics[width=\linewidth]{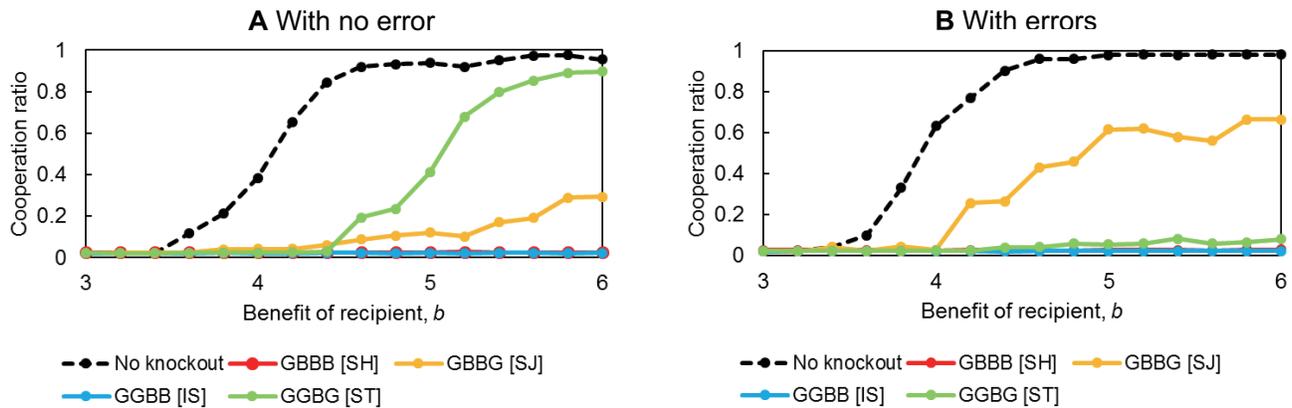}
\caption{{\bf The cooperation ratio in the norm knockout method}. Each graph shows the average cooperation ratio of 50 replications when a typical norm is knocked out. The basic parameter set is $c = 1, N = 500, R = 500, G = 1000$. To confirm the effects of errors in perception and errors in implementation, two simulations with and without error were executed. See the Supplementary Information for knockout analysis of all norms. {\bf \sf a}, The case when errors in perception ($p$) and errors in implementation ($q$) are 0. When SH or IS is knocked out, cooperation does not evolve at all. Also, when SJ, which becomes the majority for only a brief round during the process of alternation, is knocked out, cooperation evolves to the extent of only 30 percent, even when $b$ is large. Furthermore, when ST is knocked out, the range in which cooperation is achieved becomes narrow. Only when $b$ is sufficiently large can cooperation evolve. {\bf \sf b}, The case where $p = q = 0.001$. The indispensable norm is ST in addition to SH and IS. Conversely, when SJ is knocked out, cooperation evolves when $b$ is sufficiently large in the same manner as ST in {\bf \sf a}.}
\label{fig:knockout}
\end{figure}

Figure \ref{fig:knockout} shows the cooperation ratio when a particular norm is knocked out. Regardless of whether there is an error, if SH or IS is knocked out, cooperat

ion does not evolve at all. We define indispensable norms in the evolution of cooperation as the norms that, when knocked out, have an average cooperation ratio of less than 0.1 after 1,000 generations. In an environment with no errors, SH and IS are indispensable norms. In an environment with errors, SH, IS, and ST are indispensable norms.

\begin{figure}[ht]
\centering
\includegraphics[width=\linewidth]{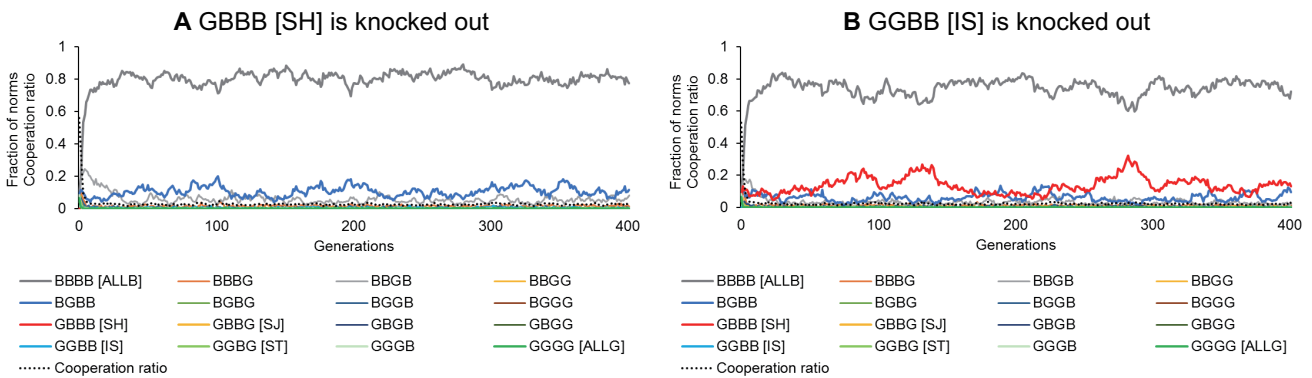}
\caption{{\bf Time series of typical simulation runs in norm knockout method}. The parameters are $b = 5,c = 1,N = 500,R = 500,p = q = 0$. {\bf \sf a}, When SH is knocked out, the strategy to eliminate ALLB does not exist. {\bf \sf b}, When IS is knocked out, SH exists only in a small population and cannot gain superiority over ALLB.}
\label{fig:knockout_time}
\end{figure}

When an indispensable norm is knocked out, cooperation does not evolve.
When cooperation evolves, alternation from strict norms to tolerant norms was observed, as shown in Figs. \ref{fig:alternation1} and \ref{fig:alternation2}.
To analyse whether alternation also occurs when a norm is knocked out, the population ratio of norms when typical norms are knocked out is displayed as time-series graphs (see Fig. \ref{fig:knockout_time}). Figure \ref{fig:knockout_time} shows the results in the cases where SH or IS were knocked out. We discovered that the first condition for the necessary process when cooperation evolves is whether SH can antagonize ALLB. No norm that resists the invasion of ALLB appears in a society in which SH does not exist. Also, in a society in which IS does not exist, SH cannot antagonize ALLB. We found that IS is a norm indispensable for SH to resist ALLB.

\section*{Discussion}

Our model offered two major findings on the evolution of cooperation on indirect reciprocity.
On the one hand, the most essential contribution is the discovery of indispensable norms by the norm knockout method. By using the norm knockout method, we were able to elucidate the existence of norms indispensable for the evolution of cooperation from a melting pot of norms. Regardless of the existence of errors, SH and IS were indispensable norms. In addition, in an environment with errors, ST is an indispensable norm. Interestingly, SH and IS are reconciled to the minorities after the cooperative regime emerges while they temporarily become major norms in the process of dynamics. We call such minority norms required for the evolution of cooperation “unsung hero norms”. The results clearly illustrate the two roles of norms: one to catalyse a cooperative regime and the other to maintain the regime. Norms having the GG** for the evaluation rule play the latter role.

On the other hand, we discovered alternation of norms. Recent analysis of evolutionary stability against the invasion of free riders could identify neither superiority among norms nor the process on the path to cooperation.
Among studies on indirect reciprocity, ours is the first exhaustive theoretical analysis on all possible norms, although several studies have addressed the comparison of two types of reciprocal norms \cite{uchida2010competition,uchida2010effect,matsuo2014}.
Others analyse the alternation of norms in direct reciprocity \cite{lindgren1992,Zagorsky2013,Berg2015}. 
We find the alternation of the norms and also discover the indispensable norms that are required to foster indirect reciprocity.

An empirical study \cite{Swakman2016} supports the co-existence of various norms in the cooperative regime and indicates that the ST norm plays a more important role in human cooperation than SJ, which is consistent with our simulation. This is because we show that the SJ norm cannot survive in the cooperative regime, while the ST one can. Our approach may provide deep insight on the evolution of cooperation because several norms absolutely play an essential role in order to evolve cooperation even though, on the surface, it seems as though they are not directly leading to the evolution of the cooperation.

The present work considers a single action rule (cooperate with Good, defect with Bad) to stress the role of multiple assessment rules.
However, the other papers stress the role of multiple co-existing action rules \cite{Brandt2005,Ohtsuki2007,Santos2016,Sasaki2016}.
Integrating the multiple assessment and action rules may be a useful extension of this paper.
We analyse what happens when one norm is absent from the population; however, we have not analysed all the indispensable combinations of norms yet.
Extending the norm knockout method to combinations of norms may also be a useful extension of this paper.

\section*{Methods}

In this section, we describe the details of our agent-based model that uses the norm knockout method.
The following model description follows the ODD protocol\cite{Grimm2010}.

\subsection*{Purpose}
The aim of the model is to understand the dynamics of norms during the evolution of cooperation, and to find indispensable norms without which cooperative societies could never emerge.
In particular, we reveal the effect of these indispensable norms on indirect reciprocity using a new methodology we call the "norm knockout method".
We utilize the giving game framework\cite{Sigmund2010} for simulation.

\subsection*{Entities, state variables, and scales}

The entities in the model are agents who play as donor and recipient in the giving game with no spatial structures.
The donor chooses cooperation or defection with a recipient using an image that the donor has to the recipient.
An image is either Good or Bad.
If a donor's image to a recipient is Good, the donor cooperates with the recipient.
If the image is Bad, the donor defects.
The group size of the model is $N$.
Each agent has it's own norm and a list of images to other agents.
The agent also has a probability of errors and a payoff of the game.

The norm of an agent is denoted as one of four possible "assessment combinations", and there are two possible "alleles (G/B)" at the "locus" for each of the four assessment combinations.
The first locus of the gene represents an assessment rule to an agent who cooperates with a Good recipient.
The second locus represents an assessment rule to an agent who cooperates with a Bad recipient.
The third locus represents an assessment rule to an agent who defects with a Good recipient.
The fourth locus represents an assessment rule to an agent who defects with a Bad recipient.
Incidentally, all agents evaluate themselves as Good.
For instance, ALLG always assesses others as Good, and thus, its "genotype" is GGGG using the above mentioned definition of the four loci.
Similarly, ALLB is described as BBBB, IS as GGBB, ST as GGBG, and SJ as GBBG.

Each agent has two different types of errors: one, the probability that the agent's updating of its evaluation of others, Good/Bad, is inverted (errors in perception), described as $p$, and two, the probability to perform an action differently from the one prescribed by its action rule (errors in implementation), described as $q$.
The evolution process of norms involves adopting a genetic algorithm \cite{Holland1975}.
State variables and initialization in the simulation are shown in Table \ref{tab:variable}.

\subsection*{Process overview and scheduling}

Our simulation runs throughout $G$ generations.
A generation consists of $R$ rounds.
The agents play the giving game $R$ times as donor in each generation.
At the end of a generation, they evolve their own norms using accumulated payoffs that are obtained in the generation.
One round has two phases: (A) a phase to play giving games and (B) a phase to update images.
After all agents play giving games in phase (A), all agents update their images in phase (B).
In phase (A), each agent becomes a donor and each donor randomly chooses a recipient from $N-1$ players excluding itself.
The donor chooses whether to give benefits to the recipient or not.
At that time, the action of the donor is inverted with the probability $q$.
The donor who cooperates pays cost $c$ and the recipient receives benefit $b$ $(b>c>0)$.
In phase (B), each agent (set to $i$) evaluates and updates an image to the other agent (set to $j$).
The new image to $j$ from $i$'s viewpoint depends on $j$'s action (C/D) as a donor in the last round and depends on the image to $k$ from the viewpoint of $i$ (G/B), where $k$ was a recipient of $j$ in the last round.
At that time, the image to $j$ is inverted with the probability $p$.
In the first round in a generation, $j$'s action (C/D) is regarded as random.

After $R$ rounds of the giving game are played in every generation, agents evolve their norm.
The evolution process of norms involves adopting the genetic algorithm \cite{Holland1975}.
Because each locus of norm has an independent meaning for assessment to others, the adaptive process should contain a combination of these elements rather than a string of norms.
We have modeled a process of updating norms not as a string of norms but rather as four different assessment rules, which enables the norms to be interpreted as different situations depending on the norm genotype.
The first, second, third, and fourth loci represent the assessment for pro-social behavior, tolerant behavior, anti-social behavior, and justified defection (punishment), respectively.
Each agent randomly selects two agents from $N$ agents (including itself) to become its parents.
For choosing parents, we adopt a roulette selection method.
This roulette selection sets a probability distribution of all agents as $\Pi_i=(U_i - U_{min})^2/\sum_{j}^{} (U_j - U_{min})^2$ , where $U_i$ denotes the agent $i$'s accumulated payoff in a generation given by $U_i = bW - cV$, with $W$ being the number of donations $i$ received in the generation and $V$ the number of donations $i$ gave.
$U_{min}$ means a minimum value of the accumulated payoffs among all.
Finally, each agent updates its norm using a uniform crossover technique.
With mutation rate $m$, each locus is inverted for maintaining the diversity of the norm space.

\subsection*{Design concepts}

\begin{description}
\item{\bf Basic principles}
An agent-based simulation is utilized to study indirect reciprocity.
We explore how different combinations of norms interact to produce an evolutionary progression towards cooperation.
\item{\bf Emergence}
A cooperative regime in the situation of social dilemma emerges from interactions among agents who have various social norms.
\item{\bf Adaptation}
The agents of the model play the giving game using their images to others.
The agents update their images to others using their norms every round.
They evolve their own norms using accumulated payoffs that are obtained in the generation.
A norm that can obtain a higher payoff can increase the population through the generation.
\item{\bf Objectives}
The objective of all agents is to maximize their own payoff.
To maximize payoff, they change their own norm at the end of each generation.
\item{\bf Learning}
The agents change their norms in each generation using a genetic algorithm.
The fitness of each agent is calculated from the accumulated payoff in the generation.
To select the parents of the agent, the model utilizes a roulette selection method.
\item{\bf Interaction}
The interaction between the agents is one to one interaction.
The giving game consists of the donor and the recipient.
There are no spatial structures in the society.
\item{\bf Stochasticity}
The interaction between agents is a stochastic process because interaction partners are chosen randomly from the society.
At the start of the simulation, each agent is randomly assigned a norm of all 16 norms.
\item{\bf Observation}
Three indexes are used for observation: average cooperation ratio in the society, the transition of norms with the greatest populations, and population ratio of each norm.
\end{description}

\subsection*{Initialization}

At the start of simulation, the norm of each agent is chosen randomly from all 16 possible norm combinations.
In the first round of each generation, the evaluation of all agents is initialized as Good \cite{Nowak1998b,Takahashi2006,Ohtsuki2015} and payoff of the agents is initialized as $0$.

\subsection*{Input data}

After initialization, the model does not include any external inputs, i.e., the number of agents ($N$), error ratio ($p, q$), benefit ($b$), and cost ($c$) are constant.

\subsection*{Submodels}

\subsubsection*{The norm knockout method}

The norm knockout method is implemented as follows.
When we knock out a particular norm, that norm is removed in the first round of each generation.
Concretely, if the norm of an agent evolves into a norm that is knocked out as a result of the adopting process, the norm of the agent is changed to one of the other 15 norms randomly.
In other words, the norm that is knocked out will never exist at all in the society.


\bibliography{refs_knockout}

\section*{Acknowledgements}

HY acknowledges Grant-in-Aid for Scientific Research (C) 15KT0133 and 26330387. HY also acknowledges Prof. Kurihara (UEC, Japan) for providing the computational resources. IO acknowledges Grant-in-Aid for Scientific Research (B) 16H03120. TS acknowledges the Austrian Science Fund (FWF): P27018-G11.

\section*{Author contributions statement}

HY initiated and performed the project. HY, IO, SU and TS designed the project, wrote the paper, and approved the submission. All authors reviewed the manuscript. 

\section*{Additional information}
The authors declare no competing financial interests.

\clearpage

\section*{Tables}

\begin{table}[ht]
\begin{center}
\caption{State variables and initialization in the simulation}
\label{tab:variable}
\begin{tabular}{lllr} 
\hline
Variable & Description & Type of variable & Initial value \\\hline
{\bf Agent} &  &  &  \\
Norm & Norm of agent & 16 types & chosen randomly \\
Image & Images to other agents & Binary (G/B) & G \\
Payoff & Accumulated payoff of the giving game & Real Number & 0 \\
$p$ & Errors in perception & Constant & \{0, 0.001\} \\
$q$ & Errors in implementation & Constant & \{0, 0.001\} \\\hline
{\bf Environment} &  &  &  \\
$N$ & Number of agents & Constant & 500 \\
$G$ & Generations of simulation & Constant & 1000 \\
$R$ & Times of playing giving game per generation & Constant & 500 \\
$b$ & The benefit of giving game & Constant & [3.0, 6.0] \\
$c$ & The cost of giving game & Constant & 1 \\
$m$ & The mutation ratio & Constant & 0.01 \\\hline
\end{tabular}
\end{center}
\end{table}

\clearpage

\section*{Figures}


\clearpage

\section*{Supplementary Information}

This section includes: 
\begin{itemize}
\item Supplementary Text S1 
\item Supplementary Tables S1 – S6
\item Supplementary Figure S1
\end{itemize}

\subsection*{Text S1}

\subsubsection*{The details of alternation of norms}

We present the details of alternating the majority of norms, as shown in Fig. 3 (in the Main Text). Table S1 shows the alternation with no error and Table S2 shows with errors.

\subsubsection*{The population of each norm}

We show the average population ratio at the 1,000th generation (50 replications). As shown in Table S3, persistent norms are able to be described as GG**.

\subsubsection*{The details of results of norm knockout method}

We comprehensively explore the indispensable norms by knocking out each of the 16 norms. Table S4 shows the cooperation ratio in which each norm is knocked out. Without errors, SH and IS are the indispensable norms. If SJ or ST is knocked out, their standard deviations (S.D.) are larger than other cases. The reason for this is that to knock out SJ or ST produces two contrasting results: cooperation dominant or defection dominant. When indispensable norms are knocked out, no cooperative regime appears at all, and the standard deviation has a very small value.

\subsubsection*{The analysis of the alternation of norms after cooperative regime achieved}

In this section, we have analysed the transition of norms with the greatest populations after a cooperation ratio exceeds 0.9 until the end of generations. For the sake of understanding the mechanism of co-evolution of norms and cooperation, we have focused on the duration of regime changes from defection to cooperation in the main text. Therefore, we stopped the calculation of the transition of the majority norm when ALLG becomes the majority norm in Fig. 3. Intuitively, it seems impossible to transit from a majority of ALLG to a majority of any other norm; however, it is well-known that a state where everyone cooperates indiscriminately is easily replaced by a state where everyone refuses to cooperate.

To clarify whether ALLG is a stable state and what would happen if more generations were considered, we show the transition of norms with the greatest population after a cooperative regime is achieved. The results (Tables S5 and S6, Fig. S1) show that the cooperation regime is maintained robustly and tolerant norms such as ALLG, GGGB, and ST coexist. Although ALLG forms the majority of the population, the other norms including GGGB, ST (GGBG), and IS (GGBB) protect against invasion from defective norms.

\newpage
\subsection*{Tables}
\setcounter{table}{0}

\begin{table}[ht]
\textbf{\textsf{Table S1.} The alternation patterns of dominant norms with no error correspond to the left panel of Fig. 3 (in the Main Text).}
Each row shows the transition of norms with the greatest populations in a period of 20 generations, before the cooperation ratio exceed	s 0.8, and 100 generations, after the cooperation ratio exceeds 0.8 (for a total of 120 generations). For the sake of visibility, we stop calculation when ALLG becomes the majority norm. Fifty replications are conducted. During this time, alternation in majority norms for a total of 156 times could be observed. A cooperative regime with a cooperation ratio exceeding 0.8 was achieved in 46 replications. For example, over 50 replications, the number of times the transition of greatest population followed (SH $\to$ SJ $\to$ ST $\to$ ALLG) was 31. Moreover, four (indicated by the dash) never had a cooperation ratio exceeding 0.8.
\begin{center}
\begin{tabular}{lr} \hline
Transition pattern of dominant strategies & No. \\\hline
SH $\to$ SJ $\to$ ST $\to$ ALLC  & 31 \\
SH $\to$ SJ $\to$ ST $\to$ SJ $\to$ ST $\to$ ALLC & 5 \\
$ - $ & 4 \\
SH $\to$ GBGB $\to$ GGGB $\to$ ALLC & 3 \\
SH $\to$ GBGB $\to$ SH $\to$ SJ $\to$ ST $\to$ ALLC & 2 \\
SH $\to$ IS $\to$ GGGB $\to$ ALLC & 2 \\
SH $\to$ IS $\to$ GGGB $\to$ SH $\to$ ALLD $\to$ SH $\to$ ALLD $\to$ SH $\to$ ALLD & 1 \\
SH $\to$ GGGB $\to$ ALLC & 1 \\
SH $\to$ GBGB $\to$ ST $\to$ ALLC & 1 \\\hline
\end{tabular}
\end{center}
\end{table}

\begin{table}[ht]
\textbf{\textsf{Table S2.} The alternation patterns of dominant norms with errors correspond to the right panel of Fig. 3 (in the Main Text).}
The setting of the table is the same as Table S1.
\begin{center}
\begin{tabular}{lr} \hline
Transition pattern of dominant strategies & No. \\\hline
SH $\to$ IS $\to$ GGGB $\to$ ALLC & 13\\
SH $\to$ GBGB $\to$ GGGB $\to$ ALLC & 8\\
SH $\to$ SJ $\to$ ST $\to$ ALLC & 7\\
SH $\to$ GGGB $\to$ ALLC & 6\\
SH $\to$ IS $\to$ ST $\to$ ALLC & 5\\
SH $\to$ ST $\to$ ALLC & 2\\
SH $\to$ GBGB $\to$ GGGB $\to$ SH $\to$ GBGB $\to$ GGGB $\to$ SH $\to$ IS $\to$ GGGB $\to$ ALLC & 1\\
SH $\to$ IS $\to$ ST $\to$ IS $\to$ GGGB $\to$ ALLC & 1\\
SH $\to$ GBGB $\to$ SH $\to$ IS $\to$ GGGB $\to$ ALLC & 1\\
SH $\to$ IS $\to$ GGGB & 1\\
SH $\to$ GBGB $\to$ IS $\to$ GGGB $\to$ GBGB $\to$ GGGB $\to$ ALLC & 1\\
SH $\to$ SJ $\to$ IS $\to$ ST $\to$ ALLC & 1\\
SH $\to$ GBGB $\to$ IS $\to$ GGGB $\to$ ALLC & 1\\
SH $\to$ GBGB $\to$ SH $\to$ IS $\to$ ST $\to$ ALLC & 1\\
SH $\to$ GBGB $\to$ GGGB $\to$ IS $\to$ GGGB $\to$ ALLC & 1\\\hline
\end{tabular}
\end{center}
\end{table}

\begin{table}[ht]
\textbf{\textsf{Table S3.} The average population ratio at the 1,000th generation ($b$ = 5).}
Each column shows without/with errors. All cells are obtained by averaging the results of 50 replications. These results show the population of each norm in which all norms exist (i.e., the norm knockout method is not used). The second row shows the average cooperation ratio ($C_{ratio}$) and standard deviation at the 1,000th generation. Below the fourth row is shown the population of each norm and its standard deviation. The norms described as GG** can coexist stably while any norm that is not GG** can barely exist. SH, which is an indispensable norm, also cannot survive. IS, which is included in four persistent norms GG**, is the most in minority of the four.
\begin{center}
\begin{tabular}{lrr} \hline
& $p = q = 0$ & $p = q = 0.001$ \\\hline
$c_{ratio}$ (S.D.) & 0.939 (0.187) & 0.980 (0.006) \\\hline
Norms & population (S.D) & population (S.D) \\
BBBB [ALLB] & 0.015 (0.085) & 0.000 (0.001) \\
BBBG & 0.001 (0.002) & 0.000 (0.000) \\
BBGB & 0.001 (0.003) & 0.000 (0.001) \\
BBGG & 0.000 (0.001) & 0.000 (0.001) \\
BGBB & 0.003 (0.007) & 0.002 (0.002) \\
BGBG & 0.003 (0.003) & 0.002 (0.002) \\
BGGB & 0.004 (0.004) & 0.004 (0.003) \\
BGGG & 0.005 (0.004) & 0.005 (0.003) \\
GBBB [SH] & 0.026 (0.109) & 0.002 (0.003) \\
GBBG [SJ] & 0.009 (0.006) & 0.005 (0.004) \\
GBGB & 0.020 (0.013) & 0.007 (0.005) \\
GBGG & 0.024 (0.012) & 0.012 (0.006) \\
GGBB [IS] & 0.132 (0.043) & 0.148 (0.040) \\
GGBG [ST] & 0.165 (0.073) & 0.201 (0.071) \\
GGGB & 0.271 (0.093) & 0.271 (0.079) \\
GGGG [ALLG] & 0.322 (0.090) & 0.341 (0.064) \\\hline
\end{tabular}
\end{center}
\end{table}

\begin{table}[ht]
\textbf{\textsf{Table S4.} Analysis of norm knockout method for each of the 16 norms.}
The table shows the cooperation ratio at the 1,000th generation in which each of the 16 norms is knocked out ($b$ = 5). Each value shows the average cooperation ratio from 50 replications and standard deviation. The cells in which the average cooperation ratio is less than 0.1 are shown in red. In this paper, we call these norms “indispensable norms”. With no error, SH and IS are indispensable norms. With errors, these two plus ST are indispensable norms.
\begin{center}
\begin{tabular}{lrr} \hline
&$p = q = 0$ & $p = q = 0.001$ \\
Knockouted norm & Mean (S.D.) & Mean (S.D.) \\\hline
BBBB [ALLB]& 0.816 (0.354) & 0.745 (0.399) \\
BBBG & 0.980 (0.008) & 0.979 (0.007) \\
BBGB & 0.978 (0.012) & 0.979 (0.007) \\
BBGG & 0.923 (0.226) & 0.961 (0.134) \\
BGBB & 0.920 (0.225) & 0.922 (0.225) \\
BGBG & 0.982 (0.008) & 0.977 (0.006) \\
BGGB & 0.959 (0.134) & 0.978 (0.007) \\
BGGG & 0.979 (0.011) & 0.959 (0.134) \\
GBBB [SH] & {\color{red}0.025 (0.004)} & {\color{red}0.026 (0.005)} \\
GBBG [SJ] & 0.120 (0.287) & 0.616 (0.457) \\
GBGB & 0.982 (0.007) & 0.977 (0.006) \\
GBGG & 0.941 (0.188) & 0.978 (0.006) \\
GGBB [IS] & {\color{red}0.023 (0.006)} & {\color{red}0.022 (0.004)} \\
GGBG [ST] & 0.412 (0.432) & {\color{red}0.055 (0.060)} \\
GGGB & 0.915 (0.225) & 0.961 (0.010) \\
GGGG [ALLG] & 0.897 (0.179) & 0.371 (0.431) \\
Without Knockout & 0.939 (0.187) & 0.980 (0.006) \\\hline
\end{tabular}
\end{center}
\end{table}

\begin{table}[ht]
\textbf{\textsf{Table S5.} The number of transitions of norms with the greatest populations with no error after a cooperation ratio exceeds 0.9 until the end of generations.}
The simulation runs 50 replications. The transition is counted when the most majority norm is superseded by other norms. Parameters: $b = 5, c = 1, N = 500, R = 500, G = 1000, p = 0, q = 0$. For example, the transition from ALLG to GGGB occurs 929 times. During this time, alternation of norms with the greatest populations for a total of 2,324 times could be observed.
\begin{center}
\begin{tabular}{llr} \hline
From & To & No. \\\hline
ALLG & GGGB & 929 \\
GGGB & ALLG & 920 \\
ST & ALLG & 225 \\
ALLG & ST & 185 \\
SJ & ST & 35 \\
ST & SJ & 6 \\
IS & GGGB & 3 \\
SH & ALLB & 3 \\
ALLB & SH & 3 \\
ALLG & IS & 3 \\
IS & ST & 3 \\
ST & IS & 2 \\
GGGB & ST & 2 \\
GGGB & SH & 1 \\
SH & SJ & 1 \\
GGGB & IS & 1 \\
ST & GGGB & 1 \\
IS & ALLG & 1 \\\hline
\end{tabular}
\end{center}
\end{table}

\begin{table}[ht]
\textbf{\textsf{Table S6.} The number of transitions of norms with the greatest populations with errors after cooperation ratio exceeds 0.9 until the end of generations.}
The setting of the table is the same as Table S5. Parameters: $b = 5, c = 1, N = 500, R = 500, G = 1000, p = 0.001, q = 0.001$.
\begin{center}
\begin{tabular}{llr} \hline
From & To & No. \\\hline
GGGB & ALLG & 968 \\
ALLG & GGGB & 950 \\
ST & ALLG & 434 \\
ALLG & ST & 426 \\
GGGB & IS & 41 \\
IS & GGGB & 40 \\
ST & GGGB & 15 \\
GGGB & ST & 14 \\
IS & ST & 12 \\
ALLG & IS & 10 \\
ST & IS & 9 \\
IS & ALLG & 8 \\\hline
\end{tabular}
\end{center}
\end{table}

\clearpage

\subsection*{Figure}

\begin{figure}[ht]
\begin{center}
\includegraphics[width=\linewidth]{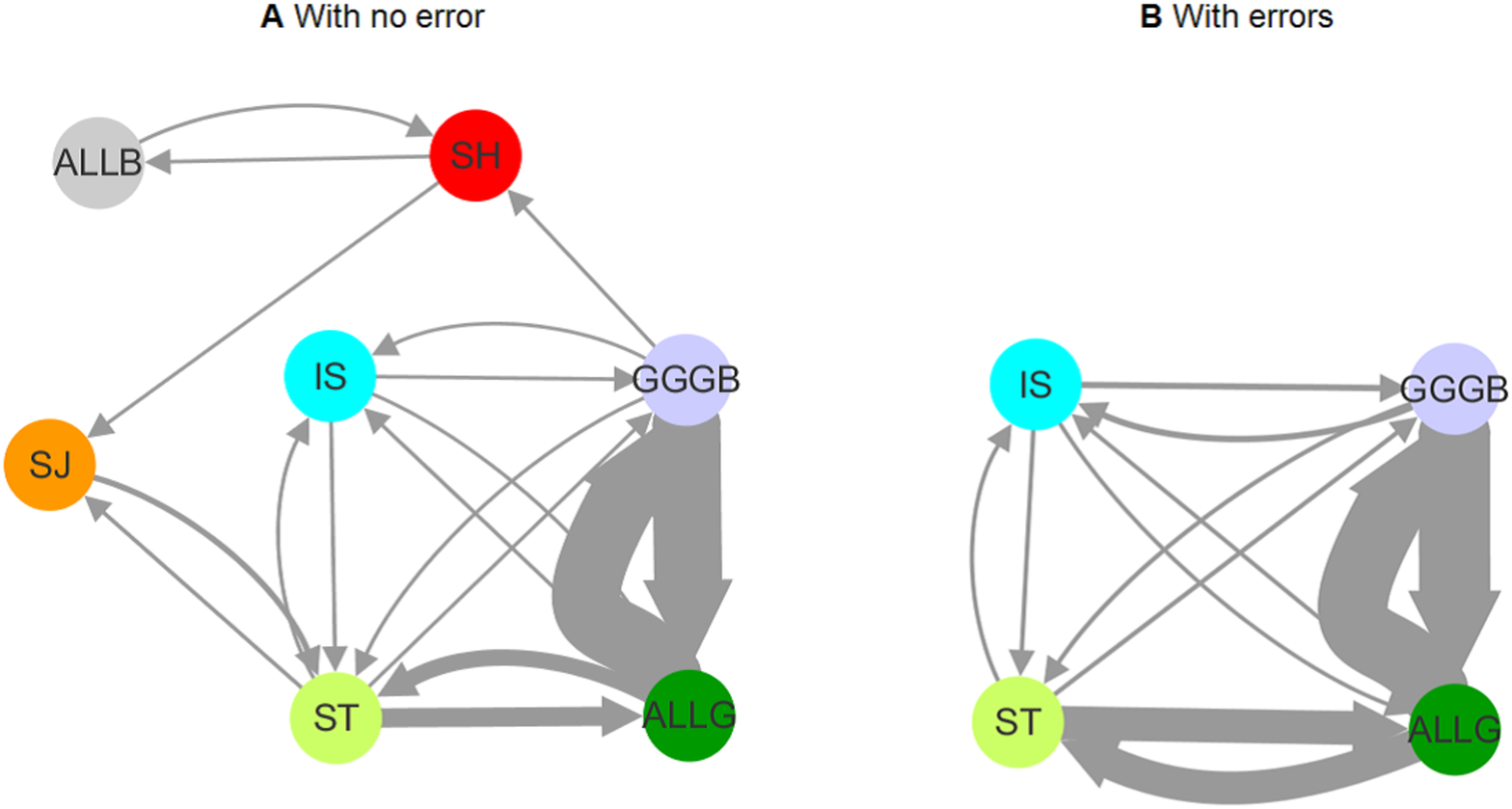}
\end{center}
\textbf{\textsf{Figure S1.} The transition diagram of norms with the greatest populations with no error (A) and with errors (B) after cooperation ratio exceeds 0.9 until the end of generations.}
Panel A is drawn using the data of Table S5 and panel B is drawn using the data of Table S6. Both panels show that the tolerant norms (such as ALLG, GGGB, and ST) coexist as the majority.
\end{figure}

\end{document}